\documentclass[12pt]{JHEP3}

\usepackage{cite}
\usepackage{amssymb}
\usepackage{amsbsy}
\usepackage{graphicx}

\newcommand{\be}{\begin{equation}} \newcommand{\ee}{\end{equation}}
\newcommand{\bea}{\begin{eqnarray}} \newcommand{\eea}{\end{eqnarray}}
\newcommand{\el}{\nonumber \\}
\newcommand{\re}[1]{(\ref{#1})}
\newcommand{\pat}{\partial}

\newcommand{\fig}[1]{Figure \ref{#1}}

\newcommand{\adot}{\dot{a}}

\newcommand{\bdot}{\dot{b}}

\newcommand{\brt}[1]{[#1]}
\newcommand{\rmd}{\textrm{d}}
\newcommand{\ack}{\acknowledgments}

\renewcommand{\H}{H_a}
\newcommand{\HH}{H_a^2}
\newcommand{\Hb}{H_b}
\newcommand{\HHb}{H_b^2}

\newcommand{\Hfour}{H_{\mathrm{4D}}}

\newcommand{\LCDM}{$\Lambda$CDM\ }
\newcommand{\GN}{G_{\mathrm{N}}}
\newcommand{\ha}{\frac{1}{2}}

\newcommand{\PRD}[1]{{\it Phys. Rev.} {\bf D#1}}

\newcommand{\PRL}[1]{{\it Phys. Rev. Lett.} {\bf #1}}

\newcommand{\NPB}[1]{{\it Nucl. Phys.} {\bf B#1}}

\newcommand{\PLB}[1]{{\it Phys. Lett.} {\bf B#1}}
\newcommand{\MNRAS}[1]{{\it Mon. Not. Roy. Astron. Soc.} {\bf #1}}
\newcommand{\APJ}[1]{{\it Astrophys. J.} {\bf #1}}

\newcommand{\JHEP}[1]{{\it JHEP} {\bf #1}}
\newcommand{\CQG}[1]{{\it Class. Quant. Grav.} {\bf #1}}

\newcommand{\AaA}[1]{{\it Astron. \& Astrophys.} {\bf #1}}

\title{Fitting oscillating string gas cosmology to supernova data}

\author{Francesc Ferrer \\
Physics Department, Washington University,
St Louis, MO 63130, USA \\
Email: \email{ferrer {\it at} physics {\it dot} wustl {\it dot} edu}}

\author{Tuomas Multam\"{a}ki \\
Department of Physics, University of Turku,
FIN-20014, Finland \\
Email: \email{tuomul {\it at} utu {\it dot} fi}}

\author{Syksy R\"{a}s\"{a}nen \\
Universit\'e de Gen\`eve, D\'epartement de Physique Th\'eorique, \\
24 quai Ernest-Ansermet, CH-1211 Gen\`eve 4, Switzerland \\
Email: \email{syksy {\it dot} rasanen {\it at} iki {\it dot} fi}}

\abstract{In string gas cosmology, extra
dimensions are stabilised by a gas of strings.
In the matter-dominated era, competition
between matter pushing the extra dimensions to expand
and the string gas pulling them back can lead to
oscillations of the extra dimensions
and acceleration in the visible dimensions.
We fit this model to supernova data, taking
into account the Big Bang Nucleosynthesis constraint
on the energy density of the string gas.
The fit to the Union set of supernova data is acceptable,
but the fit to the ESSENCE data is poor.
}


\begin{document}

\section{Introduction}

\paragraph{String gas cosmology.}

String gas cosmology is a cosmological scenario motivated
by string theory \cite{Kripfganz:1988, Brandenberger:1989, Tseytlin:1991}
(see \cite{Battefeld:2005b, Brandenberger:2008} for reviews
and \cite{Mathur:2008} for another perspective).
In the original formulation of string gas cosmology, all
spatial dimensions are treated on an equal footing: they
are all toroidal and start out at the string size.
The aim is that dynamical processes in the
early universe will allow only three dimensions to expand to
macroscopic size, while the extra dimensions are stabilised
at the string size by a gas of strings.
Assuming that the dilaton is stabilised by some other mechanism,
the string gas can stabilise the extra dimensions
during the radiation-dominated era \cite{Patil:2004, Patil:2005a}
(see also \cite{Kripfganz:1988, Watson:2003a, Berndsen:2004, Brandenberger:2005a, Patil:2005b, Berndsen:2005, Kanno:2005, Danos:2008, Sano:2008}).
However, when the universe becomes matter-dominated,
the matter will push the extra dimensions to open up
\cite{Kripfganz:1988, Patil:2004, Patil:2005a, Weiss:1986}.
It was shown in \cite{Ferrer:2005} that the gas of strings
can still prevent the extra dimensions from growing too large,
but they cannot be completely stabilised. There is a competition
between the push of matter and the pull of strings.
If the number density of the strings is too small, the extra
dimensions will grow to macroscopic size. If the strings win,
the size of the extra dimensions will undergo damped oscillations
around the self-dual radius.
The oscillations between expansion and contraction of the extra
dimensions induce oscillations in the expansion rate of
the large dimensions, which can involve alternating periods
of acceleration and deceleration \cite{Ferrer:2005}.
(This kind of a mechanism has also been studied in
\cite{Perivolaropoulos}.)

Since the oscillations can start only after the universe becomes
matter-dominated, they provide an in-built mechanism for
late-time acceleration in string gas cosmology, one that
alleviates the coincidence problem in a manner similar to
scaling and tracker fields \cite{scaling, tracker}.
The mechanism is based on ingredients already present in
string gas cosmology and does not require adding new
degrees of freedom or turning on new interactions.
However, the oscillating expansion history is quite different
from the \LCDM model which is known to be a good fit to
the observations. (For comparison of some oscillating models to
observations, see \cite{osc, Dutta:2008}.)

We compare the model studied in \cite{Ferrer:2005} to
observations of type Ia supernovae (SNe Ia), using the
Big Bang Nucleosynthesis (BBN) constraint on new radiation
degrees of freedom.
In section 2 we describe the string gas model, and in
section 3 we fit the model to the Union and ESSENCE sets of SNIa data.
The distances predicted by the model 
provide an acceptable fit to the Union data, but
the fit to the ESSENCE data is poor.
In section 4 we summarise our results
and discuss how to make the model more realistic.

\section{The string gas model}

\paragraph{The metric and the equation of motion.}

We consider the string gas model discussed in \cite{Ferrer:2005}.
The spacetime is ten-dimensional, with the metric
\bea \label{metric}
  \rmd s^2 = - \rmd t^2 + a(t)^2 \sum_{i=1}^3 \rmd x^i\rmd x^i + b(t)^2 \sum_{j=1}^6 \rmd x^j\rmd x^j \ ,
\eea

\noindent where $i=1\ldots3$ labels the visible dimensions
and $j=1\ldots6$ labels the extra dimensions.
All spatial dimensions are taken to be toroidal.
We take the value $b=1$ to correspond to extra dimensions
at the self-dual radius given by the string length
$l_s\equiv\sqrt{\alpha'}$.

We assume that the dilaton has been stabilised 
\cite{Berndsen:2004, Brandenberger:2005a, Patil:2005b, Berndsen:2005, Kanno:2005}
in a way that leaves the equation of motion of the metric
unconstrained, so that it reduces to the Einstein equation
\bea \label{eom}
  G_{\mu\nu} = \kappa^2 T_{\mu\nu} \ ,
\eea

\noindent where $G_{\mu\nu}$ is the Einstein tensor,
$\kappa^2$ is the 10-dimensional gravitational
coupling and $T_{\mu\nu}$ is the energy-momentum tensor.
(We take the cosmological constant to be zero.)

Given the symmetries of the metric \re{metric}, the energy-momentum
tensor has the form
\bea \label{emt}
  T^{\mu}_{\ \nu} = \textrm{diag}( -\rho(t), p(t), p(t), p(t), P(t), P(t),  P(t),  P(t),  P(t),  P(t) ) \ ,
\eea

\noindent where $p$ and $P$ are the pressure in the visible
dimensions and the extra dimensions,  respectively.
With \re{metric} and \re{emt}, the Einstein equation \re{eom} reads
\bea
  \label{Hubble} \kappa^2\rho &=& 3 \HH + 18 \H\Hb + 15 \HHb \\
  \label{addot} \dot\H + \HH &=& - \frac{1}{6} \kappa^2 ( \rho + 3 p ) - \frac{3}{8} \kappa^2 ( \rho - 3 p + 2 P ) + 6 \H\Hb + 10 \HHb \\
  \label{bddot} \kappa^2 ( \rho - 3 p + 2 P ) &=& 8 \dot\Hb + 24 \H\Hb + 48 \HHb \ ,
\eea

\noindent where $\H\equiv\adot/a$ is the expansion rate 
of the visible dimensions and $\Hb\equiv\bdot/b$
is the expansion rate of the extra dimensions.

\paragraph{The distance.}

In Friedmann-Robertson-Walker models, the luminosity
distance $D_L$ is determined by the Hubble rate as a function
of redshift and the spatial curvature at one time,
\bea \label{DFRW}
  D_L = (1+z) \frac{1}{\Omega_{K0} H_{a0}} \sinh\left(\Omega_{K0}H_{a0}\int_0^z \frac{\rmd z'}{H_a(z')} \right) \ ,
\eea

\noindent where $\Omega_K$ is the spatial curvature density
parameter and the subscript $0$ refers to the present day
(see e.g. \cite{Clarkson:2007, Rasanen:2008}).

The metric \re{metric} is not homogeneous and isotropic,
so the relation \re{DFRW} does not hold. The distance
should instead be calculated from the general equation
(the null geodesic shear has been neglected)
\bea \label{D}
  \pat^2_\lambda D_A &=& - \ha G_{\mu\nu} k^\mu k^\nu D_A \ ,
\eea

\noindent where $D_A=(1+z)^{-2} D_L$ is the angular diameter
distance, $\pat_\lambda$ is the derivative along the null geodesic
and $k^\mu$ is the photon momentum (see e.g. \cite{Rasanen:2008}).
We only consider light rays which propagate in the visible
directions, and \re{D} reduces to
\bea \label{DSGC}
 \H \pat_z [ (1+z)^2 \H \pat_z D_A ] &=& ( \dot{\H} + 3 \dot{\Hb} - 3 \H\Hb + 3 \Hb^2 ) D_A \ .
\eea

\noindent Only if $\Hb=0$ can we integrate \re{DSGC}
to recover (the spatially flat case of) \re{DFRW}.
The relation \re{DFRW} was formulated as a consistency
check for the FRW metric in \cite{Clarkson:2007}.
String gas cosmology provides a concrete example of a model
where the metric does not have the FRW form and the consistency
condition is violated.
In general, this is also the case in other models with
dynamical extra dimensions.
However, in models where the observers are confined to a brane, distances
along the visible directions are calculated with the induced metric
on the brane, and the evolution of the extra dimensions does not
directly enter the light propagation equation \re{D}.

\paragraph{The matter content.}

In addition to ordinary four-dimensional radiation ($\gamma$)
and pressureless matter ($m$), we have a gas of massless strings ($s$)
with winding and momentum modes in the extra dimensions
and momentum modes in the visible dimensions. The contribution
of radiation and matter to the energy-momentum tensor \re{emt} is
\bea 
  \label{gamma} && \rho_{\gamma} = \rho_{\gamma,in} a^{-4} b^{-6} \ , \quad\qquad p_{\gamma} = \frac{1}{3} \rho_{\gamma} \ , \quad\quad \ P_{\gamma} = 0 \\
  \label{dust} && \rho_{m} = \rho_{m,in} a^{-3} b^{-6} \ , \quad\quad\ \ p_{m} = 0 \ ,  \quad\quad\quad P_{m} = 0 \ ,
\eea

\noindent and for the string gas we have \cite{Ferrer:2005}
\bea
  \label{rhos} &&\rho_s = M^{-1} \rho_{s,in} a^{-3} b^{-6} \sqrt{ M^2 a^{-2} + b^{-2} + b^2 - 2 } \\
  \label{ps} && p_s = \frac{1}{3} \frac{M^2 a^{-2}}{ M^2 a^{-2} + b^{-2} + b^2 - 2 } \rho_s \\
  \label{Ps} && P_s = \frac{1}{6} \frac{ b^{-2} - b^2 }{M^2 a^{-2}  + b^{-2} + b^2 - 2 } \rho_s \ ,
\eea

\noindent where the subscript $in$ refers to the initial values,
and $M$ is the initial momentum of a string in the
visible directions in units of the string scale $l_s^{-1}$.
Note that all strings are assumed to have the same momentum.

There are four parameters in the energy-momentum tensor: the scale $M$
and the energy densities $\rho_{\gamma,in}, \rho_{m,in}$ and $\rho_{s,in}$. 
However, the parameter $M$ only determines the absolute
scale, and does not affect the dynamics, as we see by
rescaling $a\rightarrow M a$.
The evolution of the system therefore depends on
only two dimensionless combinations of the parameters,
which we take to be the following:
\bea \label{rs}
  r &\equiv& M^{-1} \frac{\rho_{\gamma,in}}{\rho_{m,in}} \el
  f_s &\equiv& \frac{\rho_{s,in}}{\rho_{\gamma,in}} \ .
\eea

\noindent Now, with the rescaled $a$, the total energy density reads
\bea \label{rho}
  \rho = \rho_{m,in} M^{-3} a^{-3} b^{-6} \left( 1 + r a^{-1} + r f_s \sqrt{ a^{-2} + b^{-2} + b^2 - 2 } \right) \ ,
\eea

\noindent and the pressures are written accordingly.

Deep in the radiation-dominated era (in particular, during BBN),
the energy density of the string gas evolves like radiation, and
contributes to the total energy density a fraction
$\Omega_{s,in} = f_s/(1+f_s)$,
given that the contribution of matter is negligible
and $b=1$ in the radiation-dominated era.
The string fraction $f_s$ is related to the effective
number of additional neutrino species $\Delta N_\nu$
by $f_s=7 \Delta N_\nu/43$ \cite{Cyburt:2004}.
From BBN we have, assuming negligible neutrino
chemical potential, the constraint
$\Delta N_\nu\leq1.5$, giving $f_s\leq0.24$,
or $\Omega_{s,in}\leq0.20$ \cite{Amsler:2008}.
Allowing for a large electron neutrino chemical potential, we have
$\Delta N_\nu\leq4.1$, which translates into
$f_s\leq0.7$, or $\Omega_{s,in}\leq0.4$ \cite{Barger:2003}.
The bound depends on the assumption that the gravitational coupling
during BBN is the same as today, which is not necessarily true
in the string gas model, since $G_N\propto b^{-6}$.
If $b<1$ today, the gravitational coupling at BBN is reduced
relative to the present value, so there is more room for new
degrees of freedom.
However, generally $b$ dips below unity
only very slightly, and typically $b>1$ today, so taking this
into account would make the constraints tighter.
It was observed in \cite{Ferrer:2005} that a requirement
for the string gas being able to keep the extra dimensions
small is $r f_s>3/2$.
There are no other constraints on $r$, since it depends on $M$,
the initial momentum of the strings in the visible directions,
on which there is no limit.

The string gas behaves like a scaling solution \cite{scaling}
in the radiation-dominated era and like a tracker
solution \cite{tracker} in the matter-dominated era
\cite{Ferrer:2005}.
The value $b=1$ is an attractor point: as long as the
initial value of $b$ is not too large ($b<\sqrt{2}$ is a
necessary condition), $b$ will rapidly evolve to unity,
and the extra dimensions are stable.
Then the energy density of the string gas behaves exactly
like radiation.
When the universe becomes matter-dominated, the string gas
starts tracking the matter as the extra dimensions expand.
When the extra dimensions are pulled back and contracted
by the strings, the visible dimensions start oscillating
between deceleration and acceleration.
(If the string gas is too weak to prevent the extra dimensions
from opening up, they will grow without bound,
and there will be no acceleration in the visible dimensions.
We are not interested in this possibility.)

\section{Comparison with observations}

\paragraph{The observations.}

We want to see how well the expansion history of the system
of equations \re{Hubble}--\re{bddot} with the energy
density \re{rho}, and the distance given by \re{DSGC},
agrees with cosmological observations.
The interpretation of many observations such
as the cosmic microwave background (CMB) and the baryon
acoustic oscillations requires perturbation theory.
The effect of the string gas and the extra dimensions on the
perturbation equations has not been completely worked out
\cite{Watson:2003b, Watson:2004, Battefeld:2005a}, so we will
consider only observations which are independent of perturbations.
Two important sets of observations which depend only on
the background are luminosity distances of SNe Ia
and the primordial abundance of light elements.
The ages of passively evolving galaxies also provide
a measurement of the Hubble parameter as a function of
redshift independent of the distance scale \cite{ages}.
In addition, there are local measurements of the Hubble
parameter, the age of the universe and the matter density.

We will use the ESSENCE SNIa dataset \cite{Davis:2007}
and the Union compilation \cite{Kowalski:2008}
separately. The Union dataset is the newest and most
comprehensive collection, but it has been analysed with
the assumption that the \LCDM model is correct.
Therefore the results cannot, strictly speaking, be used
to compare between cosmological models in an unbiased
manner, especially in the case of models which are
significantly different from $\Lambda$CDM, like the string gas model.
Therefore, we also fit to the ESSENCE dataset, which has been
analysed differently, for comparison.
We find that, despite the bias in the Union analysis,
the string gas model fits the Union dataset better than
the ESSENCE data. This could be due to more conservative
treatment of errors in the Union analysis.

We will also take into account the BBN constraint on new
radiation degrees of freedom from the observed abundance
of light elements \cite{Barger:2003}.
We will not use the data on the ages of passively evolving
galaxies, due to possible systematic effects related to stellar
evolution.
In addition to above data, a number of other general dark
energy probes have also been suggested.
In particular, the baryon acoustic oscillations \cite{bao}
and the CMB shift parameter \cite{shift} have been considered
as standard rulers. However, both of these probes suffer
from model dependence and caution should be exercised
when applying them to models other than \LCDM
\cite{shiftproblems, Komatsu:2008}.

\paragraph{The supernova datafit.}

We use a grid method to scan the model parameter space
$(r, f_s)$, because the complicated confidence contour
structure makes Monte Carlo Markov Chain methods ineffective.
We refined the grid until, for a typical size of $400\times400$,
the fit no longer improved significantly.
In order to determine the best fit values, we further
zoomed into regions with high values of $\chi^2$.
When we do not apply the BBN constraint $f_s<0.7$,
we restrict the scan to $f_s<9$, corresponding to $\Omega_{s,in}<0.9$.

\FIGURE[h,b,t]{
\includegraphics[width=\textwidth]{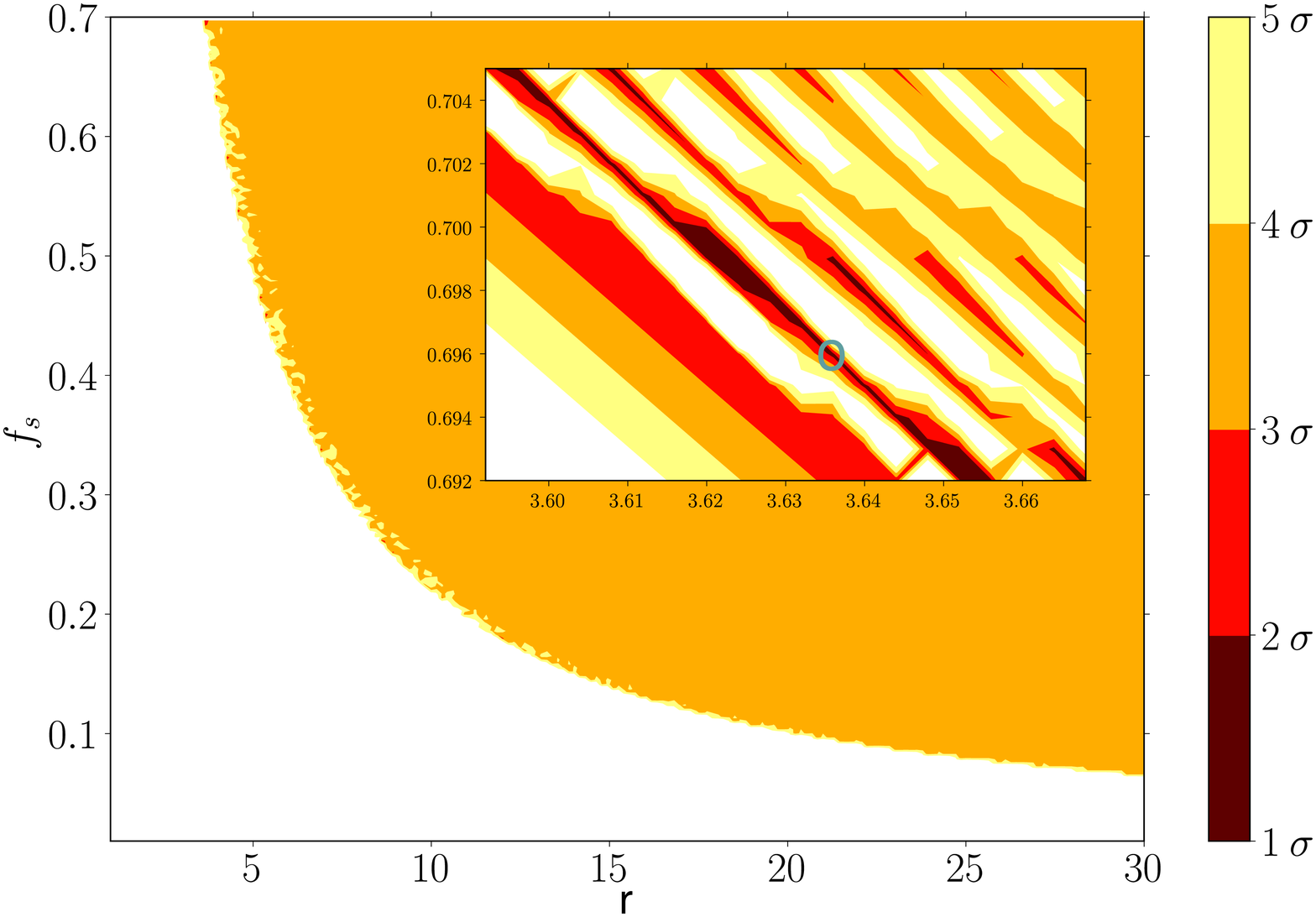}
\caption{Confidence level contours in the $(r,f_s)$-plane for the Union dataset. The best-fit model is marked with a circle.}
\label{fig:CL}}

\begin{table}
\begin{center}
\begin{tabular}{|c|ccccc|c|}
	\hline
Model &  Dataset  & $\chi_{bf}^2$ & $p (\%)$ & $r$ & $f_s$  & Remarks\\
	\hline\hline
\LCDM   &Union& 308.3& 44 & - & - & $\Omega_\Lambda=0.68$ \\
\LCDM   &Essence& 196.0& 37 & - & - & $\Omega_\Lambda=0.75$ \\
\hline
String gas & Union & 329.8 & 15 & 3.636 & 0.696 & $f_s\leq 0.7$\\ 
String gas & Union & 317.6 & 28 & 0.893 & 3.246 & $f_s\leq 9$\\ 
String gas & Essence & 262.2 & 0.03 & 3.636 & 0.696  & $f_s\leq 0.7$\\ 
String gas & Essence & 234.5 & 1 & 0.833 & 3.497 & $f_s\leq 9$\\ 
\hline
\end{tabular}
\end{center}
\caption{Goodness-of-fit and best-fit parameters for \LCDM and the string gas model for different datasets, with and without the BBN constraint.}
\label{table:chi2}
\end{table}

In \fig{fig:CL} we plot the goodness-of-fit in the
$(r,f_s)$-plane for the Union dataset
(the behaviour is similar for the ESSENCE data).
The inset shows the region around the best-fit model,
marked with a circle.
The $\chi^2$ contours have a striking structure.
The lines of equal $\chi^2$ are disjoint, and nearby points
can have radically different values of goodness-of-fit.
This is not an artifact of the analysis.
(A complicated confidence level contour structure
for an oscillating model was also found in \cite{Dutta:2008}.)
In order to have enough acceleration in the visible
dimensions at sufficiently late times, the present day
has to be in a specific location, just after the rise of
one of the first few oscillations.
The details of the oscillations, in turn, depend on
$r$ and $f_s$ in a complex manner.
(Note that the late-time evolution depends on the parameter $r$
only via the initial conditions, as the radiation term in the
energy density \re{rho} is negligible at late times.)
Also, in order to have strong acceleration, the extra dimensions
have to expand almost to the point of not turning back,
and then contract rapidly. If the extra dimensions
were to expand slightly more, they would not
turn around, and there would be no acceleration.
Therefore, the best fits are obtained on the border
of very poor fits, as seen in \fig{fig:CL}.

In Table \ref{table:chi2} we give the parameter values
as well as the $\chi^2$ (and the corresponding probability
$p$ of obtaining the data given the model) for the
best-fit string gas model with and without the BBN
constraint for both datasets; values for the best-fit
\LCDM model are shown for comparison.
Because of the complex dependence of the goodness-of-fit on
$r$ and $f_s$, we cannot definitely rule out the possibility
that there would be a better fitting model somewhere in the
regions that we did not completely study.
In the patches that we did cover in detail, the quality of fit
is already saturated at the scale visible in \fig{fig:CL} and
does not improve when zooming to smaller regions.

For the Union dataset, the $\chi^2$ for the best-fit string
gas model without the BBN constraint is 9.3 points worse
than for the \LCDM model, and 21.5 points worse when
the BBN constraint is taken into account.
For the ESSENCE data, we have $\Delta\chi^2=40.2$ and
66.2, respectively.
In \fig{fig:dl} we show the distance-redshift relation
for the best-fit model to the Union dataset with and
without the BBN constraint.
The string gas behaviour is clearly different from the
\LCDM model, and provides a worse fit to the data, though
for the Union data, the quality of the fit is still good.
The string gas model would be further disfavoured if
we took into account that it has one extra parameter
compared to $\Lambda$CDM.

\FIGURE[h,b,t]{
\includegraphics[width=\textwidth]{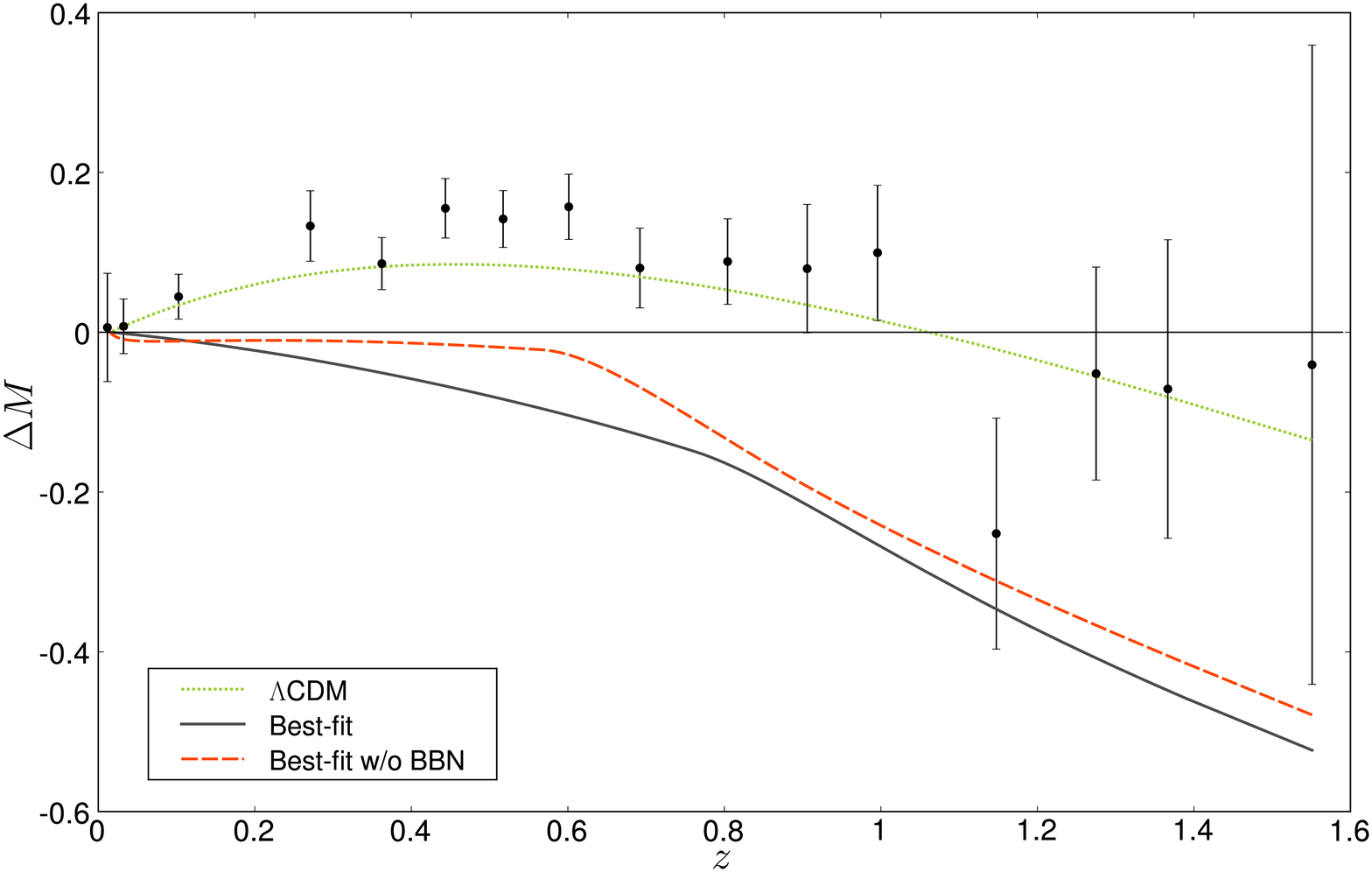}
\caption{Magnitude as a function of redshift compared to an empty universe for the Union data, for the string gas model with the BBN constraint and the \LCDM model.}
\label{fig:dl}}

\FIGURE[h,b,t]{
\includegraphics[width=\textwidth]{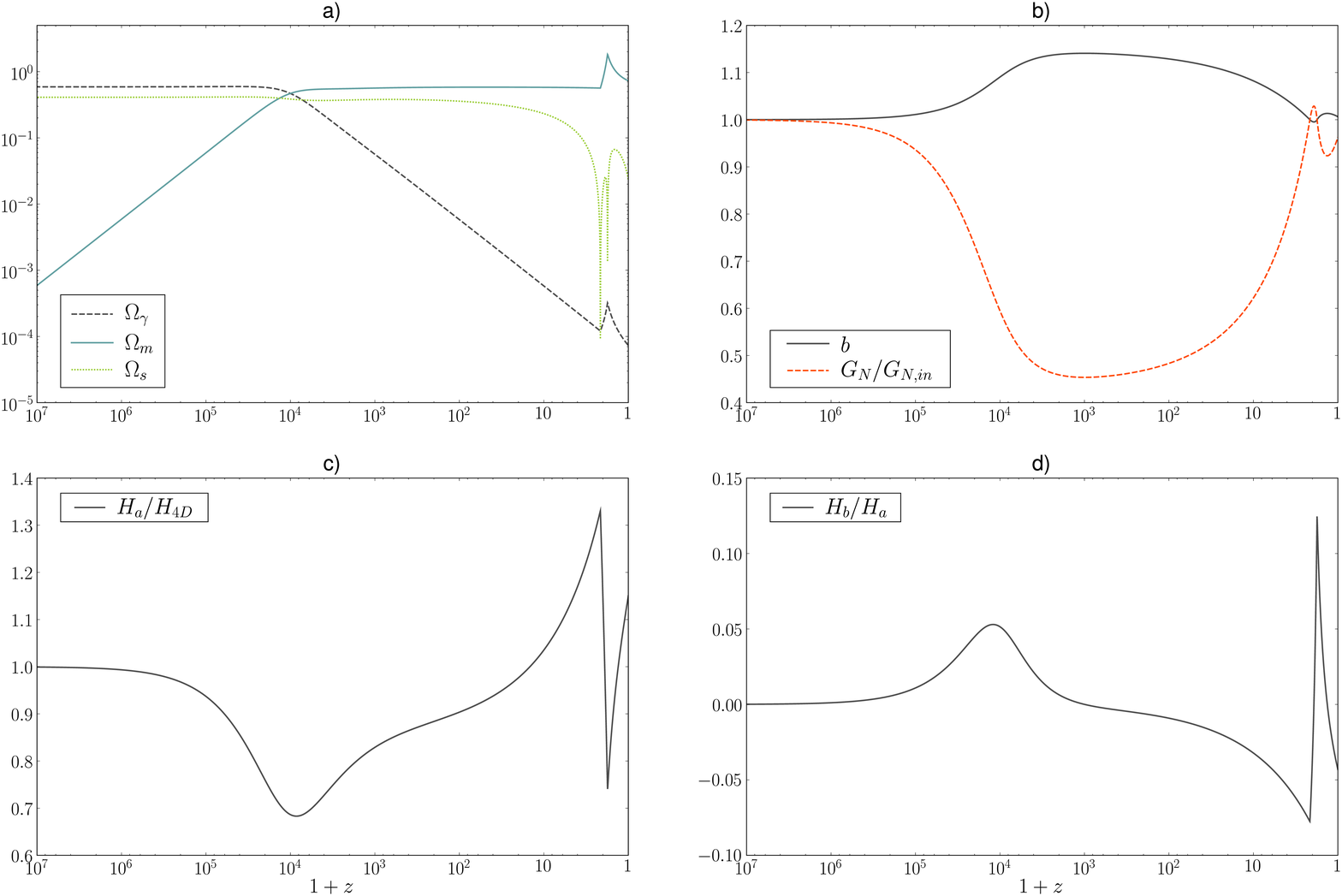}
\caption{a) Density parameters $\Omega_i \equiv \kappa^2\rho_i/(3 H_a^2)$,
b) size of the extra dimensions and Newton's constant,
c) expansion rate of the large dimensions ($H_{4D}$ is the
Hubble parameter in the usual four-dimensional case)
and d) expansion rate of the extra dimensions,
for the best-fit model to the Union data, with the BBN
constraint.}
\label{fig:bestfit}}

In \fig{fig:bestfit}, we plot some quantities for the
best-fit model to the Union dataset (with the BBN constraint
included). 
In \fig{fig:bestfit} a), we show the density parameters of radiation,
matter and the string gas. The energy density of the
string gas is completely subdominant at late times, $\Omega_{s0}=0.02$.
However, the string gas can still have a large impact on the dynamics,
because its energy-momentum tensor \re{rhos}--\re{Ps} violates the
null energy condition. When the expansion is faster than in the
Einstein-de Sitter case, the matter density parameter
$\Omega_m\equiv\kappa^2\rho_m/(3 H_a^2)$
is smaller than unity, and in principle it could be in the
observationally allowed range $\Omega_{m0}\approx$ 0.2--0.3 today.
However, for the best-fit model we have $\Omega_{m0}=0.73$,
far too large.

In \fig{fig:bestfit} b) we show the scale factor of the extra
dimensions $b$ and the four-dimensional gravitational coupling
$\GN\propto b^{-6}$. The difference between $b$ at BBN
and today is small, and well within the observational limits
discussed in \cite{Ferrer:2005}. However, $b$ deviates noticeably
from unity at last scattering at $z=1100$: $b_{LS}=1.14$,
$G_{\mathrm{N},LS}/G_{\mathrm{N},in}=0.45$. This is a generic
feature of the string gas model, because last scattering is soon after
the matter-radiation equality, when the extra dimensions start
opening up. This prediction could provide a stringent constraint.
However, quoted limits on the variation of $\GN$ (or on
new radiation degrees of freedom) from the CMB and other non-BBN
probes are model-dependent \cite{Komatsu:2008, CMB}, and rely
on perturbation theory. (Note that the
string gas does not behave like radiation at last scattering.)

In \fig{fig:bestfit} c) we show the expansion rate of the
visible dimensions $H_a$ relative to what it would
be without the extra dimensions and the string gas, denoted
by $\Hfour$. (In the matter-dominated era, $\Hfour=2/(3t)$.)
Comparing to the plot of $H_b/H_a$ in \fig{fig:bestfit} d),
we see how acceleration in the visible dimensions correlates
with contraction of the extra dimensions. 
The Hubble parameter today in the model is somewhat low, which is
related to the large value of $\Omega_{m0}$. At late times
$3\Hfour^2=\kappa^2\rho_{m,in} a^{-3}$, so we have
$\Omega_m=(H_a/\Hfour)^{-2} b^{-6}$.
In order to get enough acceleration in the recent past, it
seems that the extra dimensions must have recently collapsed,
so $b\approx1$ today. The value $\Omega_m=0.3$, for example,
then requires $H_a/\Hfour=1.8$. The maximum value of $H_a/\Hfour$
in the best-fit model is only 1.3, and the value today is 1.2.
Without the BBN constraint, the situation would be better, with
higher values of $H_a/\Hfour$.

The quantity $H_a/\Hfour$ also gives the relation between
the age of the universe and the present value of the Hubble
parameter, since $H_a/\Hfour=3 H_a t/2$ at late times.
A model-independent observational constraint on the
age of the universe is given by the ages of globular
clusters \cite{Krauss:2003}, which lead to the lower limit
$t_0\geq11.2$ Gyr at 95\% C.L. and a best-fit age of $t_0=13.4$ Gyr.
The best model-independent measure of the current
value of the Hubble parameter comes from the Hubble Key
Project \cite{Jackson:2007}. The result is sensitive to the
treatment of Cepheids, and two different analyses yield
$H_{a0}=0.73\pm0.06$ km/s/Mpc and $H_{a0}=0.62\pm0.05$ km/s/Mpc
(1$\sigma$ limits). Taking the best-fit value for $t_0$ and
the mean values for $H_{a0}$ gives $H_a/\Hfour=1.5$
and $H_a/\Hfour=1.27$, respectively. The value in the best-fit
model is too low, but not drastically so, taking into account
the uncertainties in $t_0$ and $H_{a0}$.

\paragraph{The effective equation of state.}

One reason for the poor fit is the extra-dimensional
modification of the relationship between the expansion rate
and the distance in \re{DSGC}. Rapid
oscillations of the Hubble parameter do not by themselves
rule out the expansion history. If we take the expansion rate $H_a$
for the string gas model and calculate the distance using
the FRW relation \re{DFRW}, the $\chi^2$ of the best fit
without the BBN constraint improves by 4.2 points for the
Union data and 30.7 points for the ESSENCE data, and the
fits correspond to a probability of 29\% and 20\%, respectively.

The string gas cosmology context aside, this provides an
interesting demonstration of how a model with an expansion
history radically different from \LCDM is consistent with the SNIa data.
In \fig{fig:w} a), we plot the effective equation of state $\omega_x$
of the best-fit model, defined by treating the string gas and the
extra-dimensional geometrical contributions to the Friedmann equation
as one effective component, so that \re{Hubble} reads
$3 H_a^2=8\pi G_N (\rho_{\gamma,in} a^{-4} + \rho_{m,in} a^{-3} + \rho_x)$,
with $p_x$ defined correspondingly for \re{addot},
and $w_x\equiv p_x/\rho_x$.
The variation in the effective equation of state $w_x$
is extreme: in fact, the equation of state diverges,
because $\rho_x$ passes through zero (in the plot,
we cut $w_x$ off at $\pm 2$).
The equation of state is far from constant, and far from
slowly varying, unlike assumed in most parametrisations.
(For the importance of the assumed parametrisation
of the equation of state for analysing the data, see \cite{trans}.)
Because $w_x$ diverges, the evolution of the effective
energy density is better displayed via the effective
density parameter $\Omega_x\equiv\kappa^2\rho_x/(3 H_a^2)$
shown in \fig{fig:w} b).
The effective energy density is negative for
a significant part of the evolution, as could be expected
on the basis of the strong deceleration seen in \fig{fig:bestfit} c).

\FIGURE[h,b,t]{
\includegraphics[width=\textwidth]{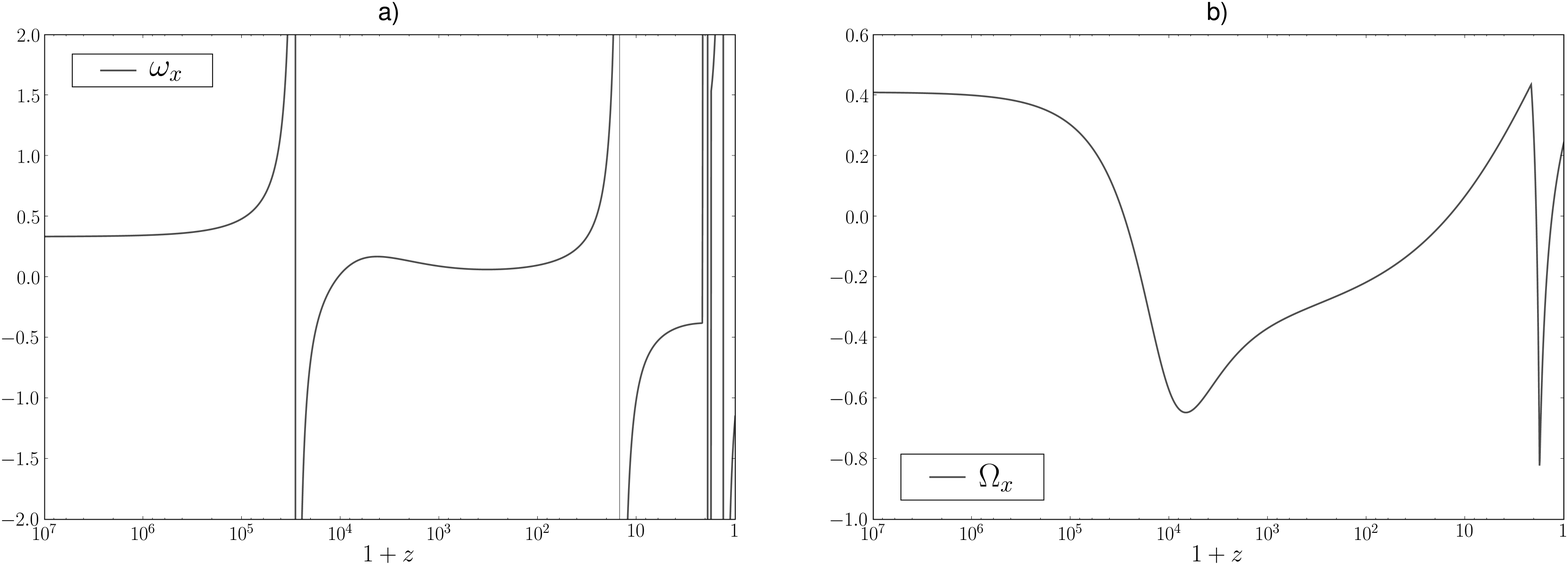}
\caption{a) Effective equation of state and b) effective
density parameter, for the best-fit model to the Union data,
with the BBN constraint.}
\label{fig:w}}

\section{Discussion} \label{sec:disc}

\paragraph{Conclusions.}

We have studied the late-time acceleration due to the
oscillations of extra dimensions in string gas
cosmology in the simple model discussed in \cite{Ferrer:2005}.
We have fitted the expansion history to the Union
and ESSENCE sets of type Ia supernovae.
The string gas model does not fit the SNIa data as
well as the \LCDM model.
With the Union SNIa data, the difference in the
goodness-of-fit is small, but the fit to the ESSENCE data
is poor.
Also, the best-fitting string gas models have a significant
fraction of the energy density during BBN in the string gas,
and taking into account the BBN constraint on new radiation
degrees of freedom makes the fit worse.
Further, we have considered a rather conservative BBN limit,
allowing for neutrino chemical potential. Since the best-fit
model is at the boundary of the region allowed by BBN,
we would expect the fit to become worse as
the BBN limit becomes more stringent.
In the model, the matter density is also too high and the
Hubble parameter today somewhat low, so taking further
observational constraints into account would be likely to
degrade the fit further.
In any case, the model can still provide a stabilisation
mechanism for the extra dimensions during the matter dominated
era, for which it was originally introduced.

Leaving aside the physical origin of the oscillations
and the constraint from BBN, the model
demonstrates how an expansion history
which is very different from the \LCDM model, with
strong oscillations of the Hubble parameter,
can still provide a good fit to the supernova data.
(In this context, it may be interesting that the Hubble parameter
inferred from observations of the ages of passively evolving
galaxies shows oscillations \cite{ages}, though
it is premature to draw strong conclusions from the data.)
The fit only becomes poor when the change in the
expansion rate-distance relationship due to the extra
dimensions is taken into account.
This in turn is a concrete example of a model where
this FRW consistency condition, discussed in \cite{Clarkson:2007},
is strongly violated.

\paragraph{Improving the model.}

As discussed in \cite{Ferrer:2005},
the energy-momentum tensor for the string gas
is expected to be more complex than \re{rhos}--\re{Ps}.
The energy density \re{rhos} corresponds to
a gas of strings which all have the same momentum
$M l_s/a$ in the visible dimensions, while
a realistic gas would have a distribution of strings
with different momenta. The evolution of
terms with different values of $M$ is qualitatively the same:
they scale like radiation in the radiation-dominated era and
start tracking the matter during the matter-dominated era
until the onset of oscillations. However, the different terms
will lead to quantitatively slightly different oscillations,
and as we have seen, the evolution is very sensitive to the
parameters of the string gas.
In order to explore this possibility, we would have to know
the distribution of string momenta, which depends on how the
string gas was created in the early universe and whether it has
thermalised.

\ack

TM gratefully acknowledges support from the Academy of Finland, project no. 111953. \\

\appendix



\end{document}